\documentclass[prl,superscriptaddress,tightenlines,showpacs,nofootinbib,notitlepage,longbibliography]{revtex4-1}
\usepackage{graphicx,amssymb,amsmath,epsf,bm}
\usepackage[normalem]{ulem}
\usepackage{stackrel}
\usepackage{soul}
\usepackage[dvipsnames]{xcolor}
\usepackage{color}
\usepackage{upgreek}
\usepackage{multirow}

\usepackage{latexsym,amsmath,amsfonts,tabularx,amssymb,bm,empheq}
\usepackage{graphicx,epstopdf}
\usepackage{xspace}
\usepackage{hyperref}
\newif\ifhyper
\hypertrue
\ifhyper
\hypersetup{
  citecolor = {green},
  colorlinks = {true}, 
  urlcolor = {blue} 
}

\makeatletter
\usepackage{epsf}
\usepackage{bm}
\usepackage[normalem]{ulem}
\usepackage{soul}
\usepackage[dvipsnames]{xcolor}
\usepackage{color}
\usepackage{upgreek}
\usepackage{multirow}
\usepackage{orcidlink}

\usepackage{latexsym}
\usepackage{amsfonts}
\usepackage{tabularx}
\usepackage{bm}
\usepackage{empheq}
\usepackage{epstopdf}
\usepackage{xspace}

\newif\ifhyper
\hypertrue
\ifhyper

\def\kbar{{\mathchar'26\mkern-9mu k}}
\def\KK{\mathcal{K}}

\makeatother

\begin{document}
\title{Exploring quantum criticality in a 4D quantum disordered system}
\date{\today}
\author{Farid Madani}
\affiliation{Univ. Lille, CNRS, UMR 8523 -- PhLAM -- Laboratoire de Physique
des Lasers Atomes et Mol\'ecules, F-59000 Lille, France}
\author{Maxime Denis}
\affiliation{Univ. Lille, CNRS, UMR 8523 -- PhLAM -- Laboratoire de Physique
des Lasers Atomes et Mol\'ecules, F-59000 Lille, France}
\author{Pascal Szriftgiser}
\affiliation{Univ. Lille, CNRS, UMR 8523 -- PhLAM -- Laboratoire de Physique
des Lasers Atomes et Mol\'ecules, F-59000 Lille, France}
\author{Jean Claude Garreau \orcidlink{0000-0003-4155-4459}}
\affiliation{Univ. Lille, CNRS, UMR 8523 -- PhLAM -- Laboratoire de Physique
des Lasers Atomes et Mol\'ecules, F-59000 Lille, France}
\author{Adam Ran\c con}
\affiliation{Univ. Lille, CNRS, UMR 8523 -- PhLAM -- Laboratoire de Physique
des Lasers Atomes et Mol\'ecules, F-59000 Lille, France}
\author{Radu Chicireanu}
\affiliation{Univ. Lille, CNRS, UMR 8523 -- PhLAM -- Laboratoire de Physique
des Lasers Atomes et Mol\'ecules, F-59000 Lille, France}

\maketitle

\textbf{Phase transitions are prevalent throughout physics, spanning thermal phenomena like water boiling to magnetic transitions in solids~\cite{HerbutBook}. They encompass cosmological phase transitions in the early universe and the transition into a quark-gluon plasma in high-energy collisions~\cite{Mazumdar2019}.  Quantum phase transitions, particularly intriguing, occur at temperatures near absolute zero and are driven by quantum fluctuations rather than thermal ones~\cite{SachdevBook}.  The strength of the fluctuations is very sensitive to the dimensionality of the physical systems, which determines the existence and nature of phase transitions. Low-dimensional systems often exhibit suppression of phase transitions, while high-dimensional systems tend to exhibit mean-field-like behavior~\cite{KardarBook}. The localization-delocalization Anderson transition stands out among quantum phase transitions, as it is thought to retain its non-mean-field character across all dimensions~\cite{Evers2008}. This work marks the first observation and characterization of the Anderson transition in four dimensions using ultracold atoms as a quantum simulator with synthetic dimensions. We characterize the universal dynamics in the vicinity of the phase transition. We measure the critical exponents describing the scale-invariant properties of the critical dynamics, which are shown to obey Wegner's scaling law~\cite{Wegner1976}. Our work is the first experimental demonstration that the Anderson transition is not mean-field in dimension four. }

\emph{This work is dedicated to the memory of Dominique
Delande, collaborator and friend, who passed away in September 2023}

Dimensionality plays a key role in physical phenomena,  especially in the context of phase transitions characterized by an order parameter, e.g. in magnets, liquid crystals, superconductors, and solids, as well as in the chiral symmetry-breaking phase transition of quantum chromodynamics and the electroweak transition in the early universe.  Quantum and thermal fluctuations increase as the dimension is lowered so that ordered phases --and thus the corresponding phase transitions-- disappear below a so-called lower critical dimension. Conversely, large dimensionality tends to increase the number of neighbors, so that mean-field theories become a good description above a certain upper critical dimension (UCD). Dimension four is the UCD for many systems, for instance magnets, superfluids, and superconductors~\cite{HerbutBook} and the paradigmatic classical Ising model~\cite{Frohlich1982,Aizenman1982,DuminilCopin}; quantum systems tend to have lower UCD~\cite{SachdevBook}, but adding disorder can increase it~\cite{Imry1975}. There are however models for which the UCD is not known, e.g. that of the Kardar-Parisi-Zhang (KPZ) equation is strongly debated~\cite{Castellano1998,Colaiori2001,Marinari2002,Fogedby2006,Oliveira2022}.

Another case where the UCD has been controversial is the Anderson transition of waves in a disordered medium \cite{Evers2008}. For dimensions larger than the lower critical dimension $d=2$\,\cite{GangOfFour}, the system presents a transition from a localized phase, where subtle destructive interferences completely suppress propagation in the medium, to a delocalized/diffusive phase where transport is possible.
The Anderson transition has been observed in three dimensions with all kinds of waves, classical or quantum, e.g. electrons in disordered electronic systems~\cite{Katsumoto1987}, cold-atom quantum-chaotic systems \cite{Chabe2008}, atomic matter waves in disordered optical potentials~\cite{Jendrzejewski2012,Semeghini2015}, acoustic \cite{Hu2008} and electromagnetic waves in random media \cite{Lahini2008,Chabanov2000,Schwartz2007}. Remarkably, all these various and very different systems are all described by the same ``universal'' functions and critical exponents, independently of microscopic details and depending only on dimensionality and the symmetries of the system, namely, in the present case, the time reversal invariance, defining the so-called \emph{orthogonal universality class}. This universal behavior is intimately related to the fact that, close to the phase transition, the physics is expected to become effectively scale invariant~\cite{Wilson:RenormGroupCriticalPhen:RMP83}.  On the theory side, Anderson transition in high dimensions is still an incredibly hard hurdle, since it is a non-standard phase transition, i.e. not captured by Landau theory in terms of an order parameter. 
While the UCD has been predicted to be four by the so-called self-consistent theory \cite{Vollhardt1982}, it has also been argued to be infinite \cite{Mirlin1994}, as also suggested by numerical simulations up to dimension six \cite{Ueoka2014,Tarquini2016}
Furthermore,  the presumed infinite UCD prevents an expansion around a mean-field-like solution, as is usually done in the context of second-order phase transitions. In particular, quantitative theoretical descriptions of the Anderson transition exist only in the vicinity of dimension two (the so-called $2+\epsilon$ expansion) and on the Bethe lattice \cite{Efetov:SupersymmetryInDisorder:97}, but there are none in dimensions greater or equal to three.

\begin{figure}
\centering \includegraphics[width=1\textwidth]{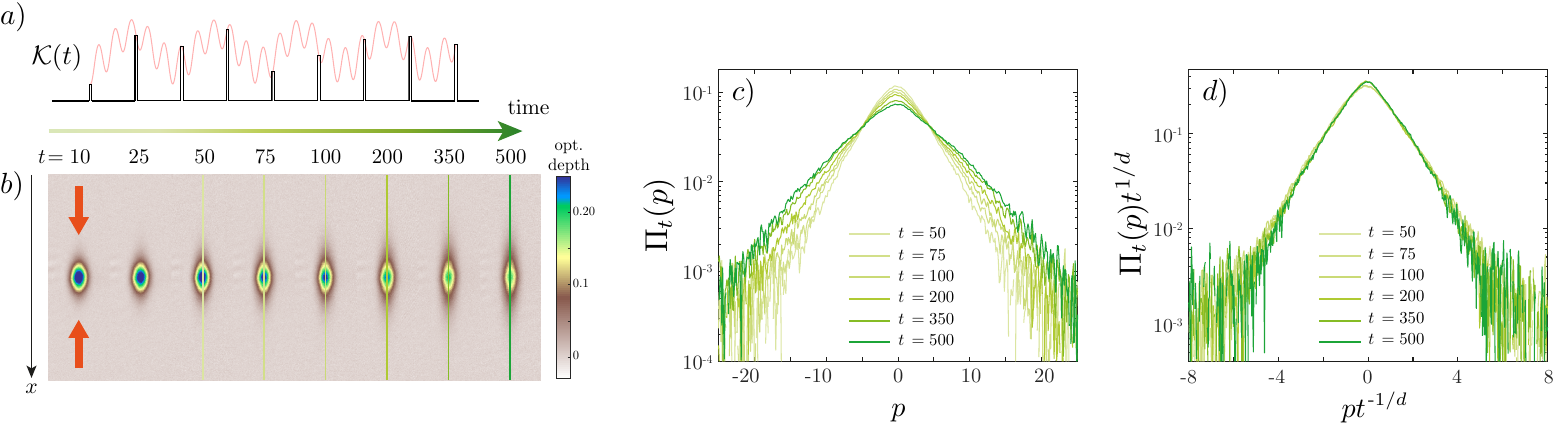} \caption{ \textbf{Experimental quantum simulation of the 4D Anderson transition.}
a) The red curve shows the kick potential's amplitude modulation envelop
$\mathcal{K}(t)$ used to synthesize
extra dimensions. The potential is pulsed periodically 
(black line).
b) Absorption images of the kicked atomic cloud at different times
at the critical point. The color scale represents the optical depth.
The sinusoidal potential is applied along the $x$ direction (red
arrows). The relevant dynamics is along the $x$ direction, which is decorrelated
from the others.
c) Momentum distributions in the $x$ direction in semi-log scale, extracted from the
absorption images {[}shown as colored lines in panel b){]}. Darker
colors represent later times. d) Scale invariance of the momentum
distribution at the critical point: after rescaling the distribution
and momenta by $t^{1/d}$ with  $d=4$,
Eq.~\eqref{eq:scale}, all curves collapse into a single one.
}
\label{fig:fancyfancy} 
\end{figure}

In this work, we experimentally realize a quantum simulation of the celebrated Anderson model~\cite{Anderson:LocAnderson:PR58} in four dimensions and study its metal-insulator (or localization-delocalization) phase transition. We find that close to the transition, the 4D scaling relations are obeyed and we determine critical exponents compatible with numerical estimates. We therefore experimentally rule out that the Anderson transition is mean-field in 4D. We also measure the full two-parameter scaling function of the momentum distribution in terms of rescaled momentum and distance to the critical point, which gives a stringent test of the scaling properties close to the Anderson transition.

Recently, various ways have been devised to create synthetic dimensions  \cite{Ozawa2019}. They can be engineered using ``parametric dimensions'' \cite{Lohse2018,Zilberberg2018} or degrees of freedom like spin \cite{Boada2012,Fangzhao2017,Yuan2018a,Mancini2015,Chalopin2020}, frequency \cite{Yuan2016,Ozawa2016}, linear \cite{An2021} and angular \cite{Luo2015} momentum, time bins \cite{Regensburger2012,Chalabi2019} or carefully designed lattice \cite{Maczewsky2020}. Here, we use a time-modulated driven gas of ultracold atoms to engineer both synthetic dimensions and disorder~\cite{Shepelyansky1987,Casati1989}. This allows us to probe experimentally, for the first time, the Anderson transition in four dimensions. 
The system is an atomic kicked rotor~\cite{Moore:AtomOpticsRealizationQKR:PRL95}
of dilute potassium atoms, submitted to a quasi-periodically modulated
pulsed laser standing wave (SW), so that only the dynamics along
the SW direction is relevant, Fig.~\ref{fig:fancyfancy}-b).
 The corresponding Hamiltonian reads 
\begin{equation}
H=\frac{p^{2}}{2}+\mathcal{K}(t)\cos(x)\sum_{n}\delta(t-n),\label{eq_H}
\end{equation}
where $x$ is the particle position (along the laser axis, measured
in units of the inverse of the kick-potential wavenumber $k_{K}$),
$p$ is the conjugate momentum, normalized such that $[x,p]=i\kbar$,
with $\kbar=\hbar k_{K}^{2}T_{1}/M$ ($M$ is the mass of the atoms,
and $T_{1}$ the kick period) and $\mathcal{K}$ is proportional
to the ratio of SW intensity to its detuning.
Time is measured in in units of $T_{1}$ and the pulse duration
is short enough to be considered instantaneous, modeled as Dirac deltas in Eq.~\eqref{eq_H}.
If the kick strength $\KK(t)$ is constant, the system is effectively 1D and displays dynamical
localization, that is, Anderson localization in
momentum space~\cite{Casati:LocDynFirst:LNP79,Moore:AtomOpticsRealizationQKR:PRL95}. Indeed, the kicks change the atoms' momentum, allowing them to ``hop'' in momentum space, while kinetic energy term give them a pseudo-random phase in momentum space, playing the role of disorder~\cite{Fishman:LocDynAnders:PRL82}. By carefully crafting the time-dependence of the $\KK(t)$, one can 
change the system's effective dimensionality  \cite{Shepelyansky1987,Casati1989}. This has allowed for a quite thorough investigation
of Anderson physics: observation of the localization in one~\cite{Moore:AtomOpticsRealizationQKR:PRL95}
and two dimensions~\cite{Manai2015}, observation of the 3D transition
\cite{Chabe2008,Lemarie2009}, characterization of its critical properties \cite{Lemarie2010,Lopez2013} and of its
universality~\cite{Lopez2012}. 

For our present purpose, following \cite{Casati1989,Chabe2008,Lemarie2009},
we realize a quasiperiodic quantum kicked rotor (QpQKR) by taking $\KK(t)=K\left(1+\varepsilon\cos(\omega_{2}t+\varphi_{2})\cos(\omega_{3}t+\varphi_{3})\cos(\omega_{4}t+\varphi_{4})\right)$, see Fig.~\ref{fig:fancyfancy}-a), with frequencies $\omega_{2}=\sqrt{5}$, $\omega_{3}=\sqrt{13}$ and $\omega_{4}=\sqrt{19}$ (the frequencies $\omega_i$, $\kbar$ and $2\pi$ must be incommensurate to avoid resonances).  This makes the driving quasiperiodic, but by defining the synthetic dimensions $x_{i}=\omega_{i}t+\varphi_{i}$ ($i=2,3,4$) and their conjugate momenta $p_{i}$ we can map this Hamiltonian onto a periodic one
\begin{equation}
\mathcal{H}=\frac{p^{2}}{2}+\omega_{2}p_{2}+\omega_{3}p_{3}+\omega_{4}p_{4}+K\left(1+\varepsilon\cos(x_{2})\cos(x_{3})\cos(x_{4})\right)\cos(x)\sum_{n}\delta(t-n).
\label{eq_4D}
\end{equation}
Here again, the kicks play the role of a hopping term of the (four-dimensional) momentum while the kinetic energy (both quadratic and linear) plays the role of disorder. One can show~\cite{Fishman:LocDynAnders:PRL82,Casati1989} that the Floquet eigenstates of the \emph{periodic} Hamiltonian $\mathcal{H}$ are eigenstates of a 4D disordered tight-binding Hamiltonian in momentum space, thus able to display an Anderson transition, see Methods. It turns out that there is no mobility edge, meaning that, for given values of $K$ and $\varepsilon$, all eigenstates are of the same nature, either localized, critical, or diffusive, making a detailed study of the critical properties possible. For $\varepsilon=0$ the system is 1D, dynamics arises in the synthetic dimensions as $\varepsilon$ increases~\cite{Lemarie:KR3DClassic:JMO10,Lopez2013}.

The experiment uses a thermal could of $\sim10^{6}$ potassium atoms (isotope $^{41}$K) prepared by evaporative cooling in a crossed optical dipole trap at a temperature $T=1.8$~$\mu$K. The cloud is kicked along the horizontal direction by a far-detuned optical standing wave created by a retro-reflected Gaussian beam with a waist of $2.8$~mm, issued from a pulsed laser at $770.229$~nm, corresponding to a detuning of $\Delta=-61.2$~GHz to the D$_{1}$ line ($770.108$~nm). The repetition frequency of the laser is $142.8$~kHz, corresponding to $\kbar=2.89$. The maximum laser power is $170$~W, and the pulse duration is 20 ns ($\sim0.3\,\%$ of the pulse period) short enough that the motion of the atoms can be neglected during the pulse and the $\delta-$kicked potential approximation is well verified. The SW amplitude modulation is generated by the RF drive of an acousto-optic modulator, thus realizing the synthetic dimensions. After a given number of kicks, the momentum distribution is measured by absorption imaging, after a time-of-flight of typically 20 ms. Throughout the experiments, atomic densities are kept below $10^{12}$~cm$^{-2}$, so that interaction effects, which might alter the characteristics of the phase transition~\cite{Cherroret:AndersonNonlinearInteractions:PRL14}, are negligible at the time scale of the kick sequence. To confirm that, we performed experiments with densities up to three times lower, without noticeable change in the measured momentum distributions.

\begin{figure}
\centering \includegraphics[width=1\textwidth]{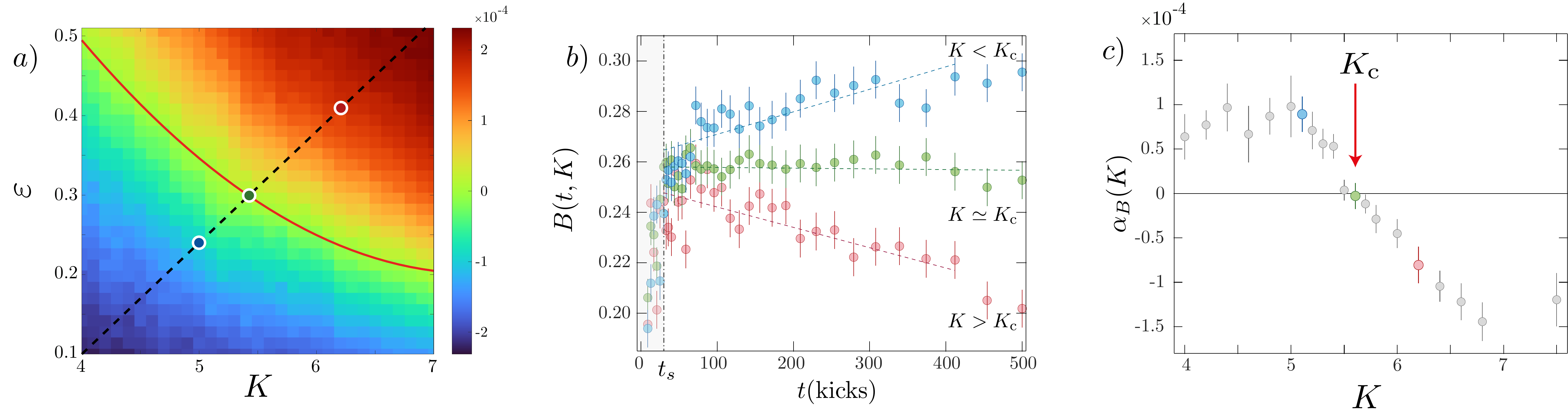} \caption{ \textbf{Locating the critical point of the Anderson transition.} 
a) Phase diagram of the QpQKR as a function of $K$ and $\varepsilon$ (numerical simulation). The color scale represents the slope of $\log \left( t^{2\alpha} \langle p^{2}\rangle_{t} \right)$, $\alpha=-1/4$, with the blue (red) region corresponding to the (de)localized phase. The critical line is in bright red along the green region. The dashed line represents the path in phase space studied experimentally. 
b) Binder parameter $B(t,K)=\Pi_{t}(0)^{2}\langle p^{2}\rangle_{t}$ in the localized ($K<K_{c}$), critical ($K\simeq K_{c}$) and diffusive ($K>K_{c}$) regimes (from top to bottom). After a short transient time ($t\lesssim t_{s}$, gray area), $B(t,K)$ is stationary at the critical point. 
c) The slope of the linear part of the Binder parameter after $t_{s}$ {[}shown as dashed lines on panel b){]} as a function of $K$. The critical point $K_{c}$ corresponds to a vanishing slope. In b) and c), the colors of the symbols match that of the corresponding dots in a).  }
\label{fig:Kc} 
\end{figure}

The phase diagram in the $(K,\varepsilon)$ space is shown in Fig.~\ref{fig:Kc}-a), see Methods, with a color code corresponding the behavior of the kinetic energy, proportional to the variance of the momentum distribution $\Pi_{t}(p)$, $\langle p^{2}\rangle_{t}=\int dp\,p^{2}\Pi_{t}(p)$. For small $K$ and $\varepsilon$ the system is localized (blue region): the atomic gas momentum distribution freezes at long times with a finite width, $\langle p^{2}\rangle_{t}\to p_{\mathrm{loc}}^{2}$ with localization length $p_{\mathrm{loc}}$ (in momentum space).  In the opposite limit, the system is delocalized/diffusive (red region): the variance of the momentum distribution increases linearly with time, $\langle p^{2}\rangle_{t}\to D\,t$, with diffusion coefficient $D$. The Anderson transition takes place on a critical line $\varepsilon_{c}(K)$ [red line in Fig.~\ref{fig:Kc}-a)], and the dynamics is universal and scale invariant in the vicinity of the transition~ \cite{GangOfFour}, see the green region in the phase diagram. This implies that the atomic momentum distribution $\Pi_{t}(p)$ obeys a universal finite-time scaling law, allowing for studying the critical behavior of the Anderson transition. The scaling law reads  
\begin{equation}
\Pi_{t}(p)=t^{\alpha}\mathcal{F}(p\,t^{\beta},\delta_K\,t^{\gamma}),
\label{eq:scale_0}
\end{equation}
with $\delta_K=K-K_c$, $\mathcal{F}$ a universal function, and critical exponents $\alpha$, $\beta$ and $\gamma$.  This universal scaling behavior holds for sufficiently long times (at short times, typically below $t_{s}=50$ kicks in the present setup, the dynamics is not universal) and close enough to the critical point. It is the analog of the finite-size scaling in standard phase transitions, with time playing the role of the finite size that the system has been able to explore.
The scaling form Eq.~\eqref{eq:scale_0} can be interpreted as follows. After a short transient dynamics ($t<t_{s}$), and as long as $\delta_K\,t^{\gamma}$ is small enough, the momentum distribution at the critical point takes a universal shape $\Pi_{t}(p)\simeq t^{\alpha}\mathcal{F}(p\,t^{\alpha},0)$, see Fig.~\ref{fig:fancyfancy}-c) and d).
If $\delta_K=0$,  it stays critical. Otherwise,  it will display this critical shape until $\delta_K t^\gamma$ is of order one, after which it will tend to a localized ($\delta_K<0)$ or diffusive ($\delta_K>0)$ shape.

\begin{figure}[t!]
\centering \includegraphics[width=1\textwidth]{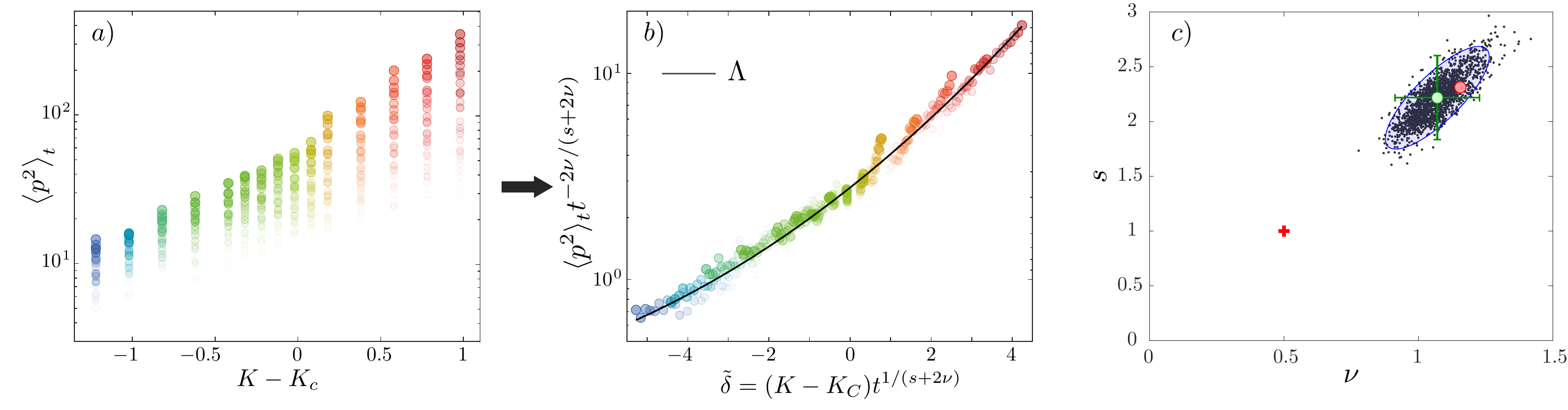} \caption{ \textbf{Determination of the critical exponents $\nu$ and $s$.} a) Kinetic energy $\langle p^{2}\rangle_{t}$ as a function of the distance to the critical point $K-K_{c}$ at various times, with $K_{c}=5.62$. Each value of $K-K_{c}$ is encoded in a different color, which goes from lighter to darker as time increases. b) Same data after rescaling, the black line corresponds
to the optimal scaling function $\Lambda(\tilde{\delta})$ using
$\nu$ and $s$ as fitting parameters, here shown at the optimal values $\nu_{\rm exp}$ and $s_{\rm exp}$, see Methods. 
c) Joint probability distribution of the exponents $\nu$ and $s$ obtained by bootstrap (see Methods). We obtain $\nu_{{\rm exp}}=1.07\pm 0.16$ and $s_{{\rm exp}}=2.22\pm0.38$ (blue ellipse corresponds to the $95\%$ confidence region, green dot with error bars corresponding to the most probable value and $95 \%$ confidence intervals).  The best numerical
values for the critical exponents for the Anderson transition in $d=4$ are $\nu_{\mathrm{num}}=1.156\pm0.014$ and $s_{\mathrm{num}}=2.312\pm0.028$ (using Wegner's scaling law), shown
as a red dot. The red cross represents the self-consistent theory prediction $\nu_{\rm sc}=1/2$ and $s_{\rm sc}=1$.
}\label{fig:nu_v_d}
\end{figure}

The exponents in Eq.~\eqref{eq:scale_0} are not independent but are related by the normalization of the momentum distribution and by the scaling of $p_{\mathrm{loc}}$ and $D$ \cite{Evers2008}:  in the localized phase ($\delta_K<0$), $p_{\mathrm{loc}}\propto|\delta_K|^{-\nu}$, defining the critical exponent $\nu$;  in the diffusive phase ($\delta_K>0$), with diffusion coefficient $D\propto\delta_K^{s}$.  As shown in the Methods, this implies $\alpha=\beta=-\nu/(s+2\nu)$ and $\gamma=1/(s+2\nu)$. Moreover, the exponents $s$ and $\nu$ have been predicted to obey Wegner's scaling law $s=(d-2)\nu$ for $d$-dimensional systems~\cite{Wegner1976}. 
It has been verified in three-dimensional electronic systems \cite{Itoh2004}, although with critical exponents in disagreement with the best numerical estimates \cite{Ueoka2014} and state-of-the-art cold atoms experiments \cite{Chabe2008}.
Studying the critical behavior Eq.~\eqref{eq:scale_0} for the 4D Anderson transition allows us to experimentally determine the scaling exponents $\nu$ and $s$, and test Wegner's scaling law.

We experimentally studied the momentum distribution across the Anderson transition by following the path in the phase diagram indicated as a dashed black line in Fig.~\ref{fig:Kc}-a). Our first goal is to locate the critical kick strength $K_{c}$; for that, we take advantage of the above-mentioned scale invariance of the distribution shape at the critical point. For this task, it is convenient to introduce a Binder-like parameter that does not scale at the transition~\cite{Binder1981}. We observe that on the one hand, $\langle p^{2}\rangle_{t}=t^{-2\alpha}\Lambda(\delta_K\,t^{\gamma})$, with $\Lambda(\tilde{\delta})=\int d\tilde{p}\,\tilde{p}^{2}\mathcal{F}(\tilde{p},\tilde{\delta})$ a universal scaling function, with $\tilde{\delta}=\delta_K t^{\gamma}$ and $\tilde{p}=pt^{\alpha}$. On the other hand, we have $\Pi_{t}(0)^{2}=t^{\alpha}\mathcal{F}(0,\tilde{\delta})$. We thus define $B(t,K)=\Pi_{t}(0)^{2}\langle p^{2}\rangle_{t}=b(\delta_K\,t^{\gamma})$ with $b(\tilde{\delta})=\mathcal{F}(0,\tilde{\delta})^{2}\Lambda(\tilde{\delta})$ a universal scaling function. Being the product of the distribution's amplitude squared by its second moment, this Binder parameter is a constant if the shape of the distribution remains unchanged during its time evolution. The critical point $K_{c}$ can thus be identified as the value of $K$ at which the Binder parameter is constant, while for other values of $K$ it evolves towards the asymptotic localized (exponential) or diffusive (Gaussian) shapes. Importantly, using the Binder parameter allows us to locate the critical point without any prior assumption on the critical scaling (namely, here, on the critical exponents $\nu$ and $s$).

Figure~\ref{fig:Kc}-b) displays the Binder parameter calculated from the experimental data for three values of $K$. The curve in green dots is almost horizontal after a time $t>t_{s}=50$, the Binder parameter is about constant, signaling the proximity with the critical point. The curves in blue and red dots vary almost linearly with time, indicating, respectively, a value of $K$ smaller or larger than $K_{c}$. Hence, we can use the slope of such curves as an indicator of the distance to the critical point. Adding more values of $K$ we obtain the plot displayed in the panel c), from which we can extract the value $K_{c}\simeq5.62\pm0.15$, in good agreement with numerical simulations $K_{c}\simeq5.45\pm0.05$ (error bars correspond to one standard deviation). 

\begin{figure}[t!]
\centering \includegraphics[width=0.95\textwidth]{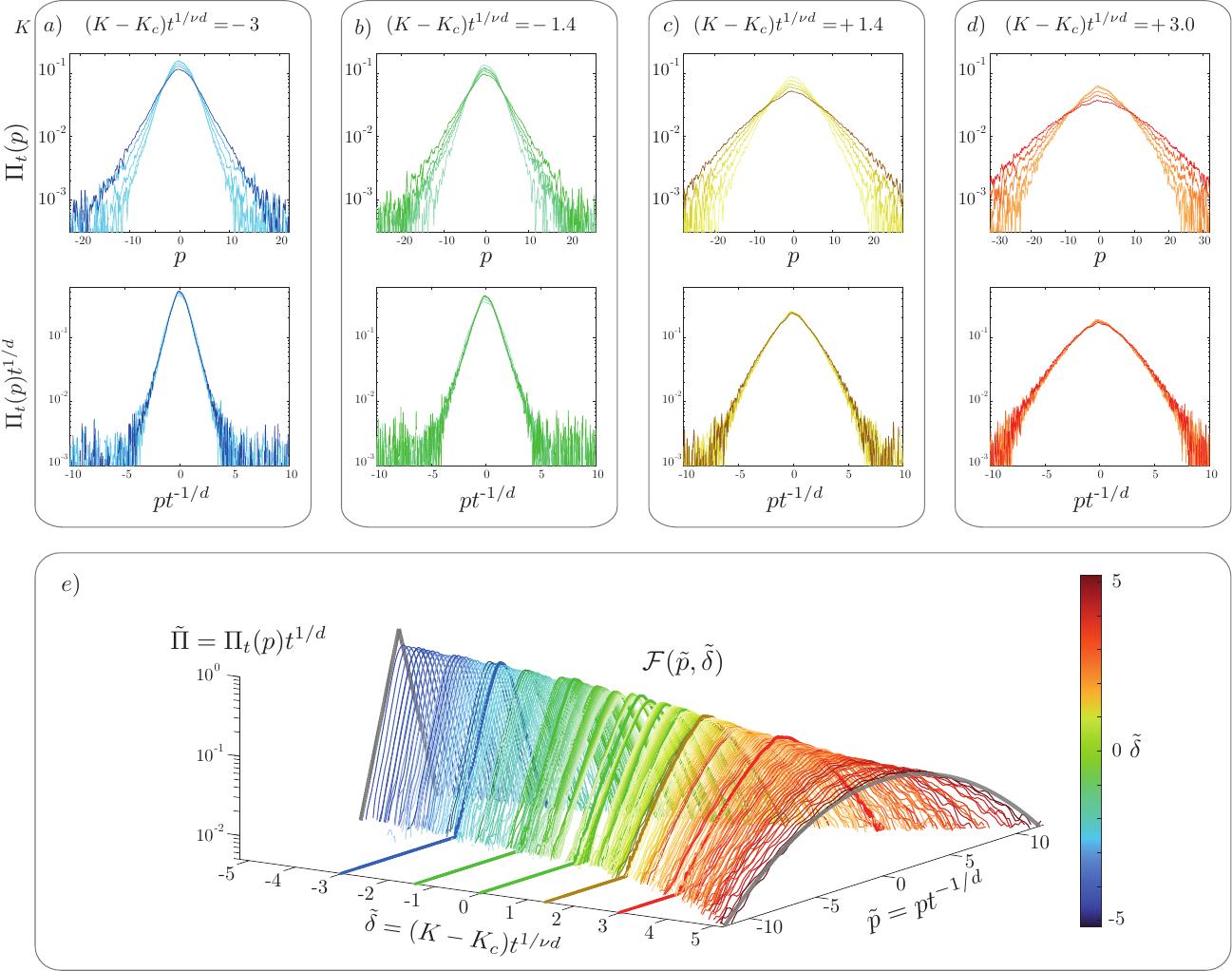} 
\caption{ \textbf{Two-parameter scale invariance near criticality of the 4D
Anderson transition.} 
Top panels: raw (top) and rescaled (bottom) momentum distributions, averaged over 20 experimental realizations, in the localized $\delta_{K}<0$ {[}a) and b){]} and diffusive ($\delta_{K}>0$) {[}c) and d){]} regimes. Colors encode the value of the scaling parameter $\tilde{\delta}$, darker tones representing longer times. 
e) Reconstruction of the two-parameter scaling function. Data obtained for various values of $K$ and $t$, properly rescaled, all fall on a smooth surface corresponding to the scaling function $\mathcal{F}(\tilde{p},\tilde{\delta})$. The data of the bottom plot in panels a), b), c), and d), as well as the data at the critical point [Fig.~\ref{fig:fancyfancy}-d)], are shown as thick lines.}
\label{fig:scale} 
\end{figure}

After finding the location of the critical point, we determine the critical exponents $\nu$ and $s$, see Methods for details. Fig.~\ref{fig:nu_v_d}-a) shows the values of the kinetic energy $\langle p^{2}\rangle_{t}$, measured at various times as a function of $\delta_K$. Rescaling this data with trial values of the exponents allows us to optimize the collapse of $\langle p^{2}\rangle_{t}$ into a single curve, see  Fig.~\ref{fig:nu_v_d}-b), using a least-square \cite{Zhang2012}.
To assert the uncertainty of the critical exponents, we use a bootstrap method \cite{Dogra2023}, which gives the joint probability distribution of the critical exponents shown in Fig.~\ref{fig:nu_v_d}-c).  The optimal values of the exponents are  $\nu_{{\rm exp}}=1.07\pm 0.16$ and $s_{{\rm exp}}=2.22\pm0.38$ (error bars correspond to $95 \%$ confidence intervals), with Wegner's scaling law $s=(d-2)\nu$ well satisfied for $d=4$.
This is in good agreement with the best numerical estimate $\nu_{\mathrm{num}}=1.156\pm0.014$ and $s_{\mathrm{num}}=2.312\pm0.028$ (using Wegner's scaling law) for a dimension four Anderson model~\cite{Ueoka2014}, marked as a red dot. The so-called self-consistent theory \cite{Vollhardt1982} predicts a UCD equal to four, with $\nu_{\rm sc}=1/2$ and $s_{\rm sc}=1$ for any $d\geq4$, marked as a red cross, while numerical simulations indicate non-trivial $\nu>1/2$ at least up to $d=6$~\cite{Ueoka2014,Tarquini2016}.  Our measurement is about $7$ standard deviations away from the mean-field value $\nu=1/2$, and thus constitutes the first experimental demonstration that $d=4$ is \emph{not} the Anderson transition's upper critical dimension.

We now verify experimentally the two-parameter scaling law Eq.~\eqref{eq:scale_0}, which we find convenient to reexpress as 
\begin{equation}
\Pi_{t}(p)=t^{-1/d}\mathcal{F}(p\,t^{-1/d},\delta_K\,t^{1/\nu d}),
\label{eq:scale}
\end{equation}
where we use Wegner's scaling law $s=(d-2)\nu$, with $d=4$ and $\nu=1.156$.
We have shown in Fig.~\ref{fig:fancyfancy}-d) that, at criticality $K=K_{c}$, the momentum distribution is indeed scale-invariant. This property also holds for $K\neq K_{c}$ provided $\tilde{\delta}=\delta_{K}t^{1/\nu d}$ is kept constant. This is shown in the top plots of Figs.~\ref{fig:scale}-a) and \ref{fig:scale}-b) on the localized side of the transition $\tilde{\delta}<0$ and in the top plots of Figs.~\ref{fig:scale}-c) and \ref{fig:scale}-d) on the diffusive side $\tilde{\delta}>0$ (the corresponding values of $t$ and $\delta_K$ are given in Methods). Having tested the two-parameter scale invariance, we can experimentally reconstruct the shape of the scaling function $\mathcal{F}(\tilde{p},\tilde{\delta})$. While one-parameter scaling functions at criticality have been measured experimentally, e.g. for the 3D Ising \cite{Damay1998,Bonetti2000} and 1D KPZ \cite{Fontaine2022} universality classes, here we measured a \emph{two-parameter} scaling function, which fully characterizes the dynamics close to the critical point. For that, we applied the scaling procedure to all the data displayed in Figs.~\ref{fig:Kc} and \ref{fig:nu_v_d}; the result is shown in Fig.~\ref{fig:scale}-e).
We observe that the data form a smooth surface corresponding to $\mathcal{F}(\tilde{p},\tilde{\delta})$, evolving from a localized, exponential shape for $\tilde{\delta}<0$ to a delocalized, Gaussian shape for $\tilde{\delta}>0$.

In conclusion, we presented the first experimental study of a quantum phase transition in four spatial dimensions, three of them being synthetic dimensions added to a physically 1D system. The critical scaling properties of the Anderson localization-delocalization transition were explicitly exhibited, including the full two-parameter scaling function, and the resulting critical exponents were found in good agreement with the numerical simulations of the Anderson model (\emph{not} that of the QpQKR). These results also show that $d=4$ is not the upper critical dimension and thus may serve as a benchmark for the still missing quantitative theory of the Anderson transition. The techniques used here might be extended to higher dimensions and other kinds of systems, thus opening new ways to test theories with varying dimensionality.

\vspace{0.5cm}
\emph{Acknowledgements:} We thank C. Cherfan for his contributions to the early stages of the experiment, and V. Vuatelet for preliminary numerical simulations on the problem. We thank C. Hainaut for useful discussions, and A. Amo, C. Hainaut, and G. Lemari\'e for a careful reading of the manuscript. This work was supported by the Agence Nationale de la Recherche (ANR) through Research Grants MANYLOK No. ANR-18-CE30-0017 and Labex CEMPI (GrantNo. ANR-11-LABX-0007-01), by CPER Wavetech, and also by the PHC Cogito and CNRS IEA programs. The Contrat de Plan Etat-Region (CPER) WaveTech is supported by the Ministry of Higher Education and Research, the Hauts-de-France Regional council, the Lille European Metropolis (MEL), the Institute of Physics of the French National Centre for Scientific Research (CNRS) and the European Regional Development Fund (ERDF).

\bibliography{addt,bibli,artdatabase_book}

\section{Methods}

\textbf{Measurement procedure and experimental uncertainties.} The average kinetic energy is determined by fitting the quantity $\Pi_t(p)\times p^2$ with a functional form (Lobkis-Weaver distribution), following~\cite{Hainaut:CFS:NCM18}. Experiments, corresponding to a given kick number and $(K,\varepsilon)$ couple, were typically repeated and averaged between three and six times. The relative dispersion of $\langle p^2 \rangle_t$ has a standard deviation of approximately $10\%$ in the experiments, and is almost independent of the experimental parameters. The precise control of the kick strength is also crucial for measuring the critical exponents. The kick laser power is controlled and modulated with an acousto-optical modulator, whose transfer function was regularly remeasured (typically every ten experimental cycles) using the a beam pick-off on a photodiode, to ensure long-term stability.  While the QpQKR experiments were performed with a thermal cloud and at low density, we used a $^{41}$K BEC in order to precisely determine the value of the kick strength. This was achieved by pulsing the lattice beams and observing an atom diffraction pattern in time-of-flight vs. the power of the lattice beams. We estimate that the uncertainty of the measurement of the kick amplitude strength, as well as the achieved long-term stability, are on the order of $1 \%$.\\

\textbf{Mapping on a disordered model.}
Following \cite{Lemarie2009}, the dynamics of the system can be mapped onto an effectively periodic one by introducing three extra dimensions, represented by canonically conjugated position and momentum operators $x_{i}$ and $p_{i}$, $i=2,3,4$. Defining the effective Hamiltonian $\mathcal{H}=\frac{p^{2}}{2}+\omega_{2}p_{2}+\omega_{3}p_{3}+\omega_{4}p_{4}+K\left(1+\varepsilon\cos(x_{2})\cos(x_{3})\cos(x_{4})\right)\cos(x)\sum_{n}\delta(t-n).$ The effective Hamiltonian is thus four-dimensional, with linear kinetic energy for the extra dimensions. Furthermore, the four-dimensional wavefunction with a ``plane source'' initial condition $\Psi(x,x_{2},x_{3},x_{4})=\psi(x)\delta(x_{2})\delta(x_{3})\delta(x_{4})$ has the same dynamics as that of the one-dimensional quasiperiodic system with initial condition $\psi(x)$.
The Floquet eigenstates $\Phi$ of the periodic Hamiltonian $\mathcal H$ with eigenvalue $e^{i\epsilon/\kbar}$ are eigenstates of a 4D tight binding model $\epsilon_{\mathbf{p}}\Phi_{\mathbf{p}}+\sum_{\mathbf{q}}t_{\mathbf{q}}\Phi_{\mathbf{p}-\mathbf{q}}=0$,  where $\mathbf{p}=(p,p_2,p_3,p_4)$ and $\mathbf{q}$ are the four-dimensional momenta on a hypercube. 
The on-site energies are $\epsilon_{\mathbf{p}}=\tan\left[\frac1{2\kbar}\left(\epsilon-\frac{p^2}{2}-\omega_2 p_2-\omega_3 p_3-\omega_4 p_4\right)\right]$ and the short-range hopping amplitude $t_\mathbf{q}$ is the four-fold Fourier transform of $\tan\left(\frac{K}{2\kbar} (1+\varepsilon \cos(x_2)\cos(x_3)\cos(x_4) ) \cos(x)\right)$. If $\kbar$ is incommensurate with $2\pi$, $\epsilon_{\mathbf{p}}$ is a pseudo-random sequence, and the tight-binding model is disordered Anderson model in momentum space  \cite{Lemarie2009}. Note that since the effective Hamiltonian is periodic in $x$ with period $2\pi$, momentum $p$ can be split into a quasi-momentum $\beta \kbar$, $\beta \in [0.5,0.5]$, and an ``integer'' part $n\kbar$, $n\in \mathbb{Z}$. The quasi-momentum $\beta$ is conserved during the dynamics and plays the role of a disorder realization in $\epsilon_{\bf p}$. Furthermore the sum $\sum_{\mathbf{q}}$ above is only over the integer part of the momentum.
\\

{\bf Numerical determination of the phase diagram and $K_c$.} The phase diagram and the location of the critical point are determined numerically by analysing the variation of the kinetic energy's scaling function $\Lambda=\langle p^2\rangle_t t^{-2/d}$. We performed numerical simulations of the QpQKR model and computed the time evolution of $\langle p^2\rangle_t$ for different values of the $(K,\varepsilon)$ couple. The results were obtained by averaging the momentum distributions corresponding to $10^4$ values of the quasi-momentum $\beta$, uniformly distributed in $[-0.5,0.5]$, and kick numbers up to 5000. The phase diagram was obtained by computing the slope of $\Lambda=\langle p^2\rangle_t t^{-1/2}$ vs. $pt^{-1/4}$ (corresponding to $d=4$). The critical region in Fig.~\ref{fig:Kc} corresponds to the $(\epsilon,K)$ couples for which $\Lambda$ is constant, yielding a zero-slope. The intersection between the critical line and the $(K,\varepsilon)$ path used in the experiment is obtained at $K=K_c=5.45\pm0.05$.\\

\textbf{Scaling form and critical exponents.}
From Eq.~\eqref{eq:scale_0}, the normalization of the momentum distribution $\int dp\, \Pi_t(p)=1$ for all $\delta_K$ and all times implies $\alpha=\beta$. Furthermore, this implies $\langle p^2\rangle_t = t^{-2\alpha}\Lambda(\delta_K t^\gamma)$. On the one hand, in the localized phase ($\delta_K<0$), $\langle p^2\rangle_t$ will become time independent at very long times, $\lim_{t\to\infty}\langle p^2\rangle_t=p_{\rm loc}^2$, defining the localization length $p_{\rm loc}$. It diverges at the transition as $p_{\rm loc}\propto |\delta_K|^{-\nu}$, defining the exponent $\nu$. This implies $\Lambda(\tilde\delta) \propto |\tilde\delta|^{2\alpha/\gamma}$ for $\tilde\delta<0$ and $|\tilde\delta|\gg 1$, and thus $2\alpha/\gamma=-2\nu$. On the other hand, in the diffusive phase ($\delta_K>0$) $\langle p^2\rangle_t$ grows linearly in time at long times, $\lim_{t\to\infty}\langle p^2\rangle_t=D t$. The diffusion coefficient $D$ vanishes as $\delta_K^s$ close to the transition, defining the critical exponent $s$. This implies $\Lambda(\tilde\delta)\propto \tilde\delta^s$ for $\delta_K\gg 1$ and $s\gamma -2\alpha=1$. Finally, assuming Wegner's scaling law, which relates the exponents $s$ and $\nu$ as $s=(d-2)\nu$, we would find $\alpha=-1/d$ and $\gamma=1/\nu d$. Our experimental measurements of the critical exponents ($\nu_{\rm exp}=1.07 \pm 0.16$ and $s_{\rm exp}=2.22 \pm 0.38$, see text) are in good agreement with Wegner's law for $d=4$.\\ 

\begin{figure}[t!]
\centering \includegraphics[width=0.95\textwidth]{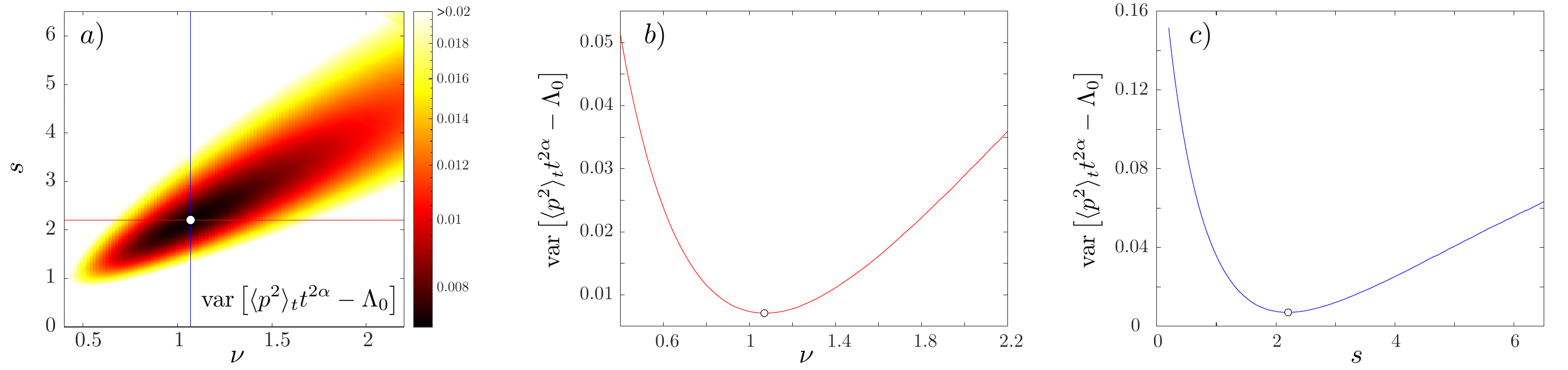} \caption{ \textbf{Determination of the critical
exponents} a) Color plot showing the variance of the dispersion of the scaled kinetic energy around $\Lambda_0$, as a function of the critical exponents $\nu$ and $s$. The minimum of this variance corresponds to the white point. The corresponding cross sections are shown in panels b) and c).}
\label{fig:variance} 
\end{figure}

\textbf{Fitting procedure for $\nu$ and $s$:} Our optimal scaling procedure uses the exponents $\nu$ and $s$ as fit parameters and consists in searching for the best collapse of the kinetic energy $\langle p^2\rangle_t$ into a single scaling function $\Lambda$, for all values of $(K,\varepsilon)$ and $t$. We use a set of values of $\langle p^2 \rangle_t$ obtained along the $(K,\varepsilon)$ path shown in Fig.~\ref{fig:Kc}.a), in the vicinity of $K_c$ (typically $|K-K_c|\leq 1$), and at times $t>30$ kicks. We use trial values of the two critical exponents $\nu$ and $s$ to compute the two scaled quantities $t^{2 \alpha}\langle p^2\rangle_t $ and $\tilde\delta=\delta_K t^\gamma$, with $\alpha=-\nu/(s+2\nu)$ and $\gamma=1/(s+2\nu)$, see text. We then compute an average curve $\Lambda_0(\tilde\delta;\nu,s)$, using a sliding average with a window of typically 1/20 of the total range of $\tilde \delta$. We interpolate this average curve and calculate the total variance of the quantity $\langle p^2\rangle_t t^{2\alpha}-\Lambda_0(\tilde\delta;\nu,s)$. The result is shown in Fig.~\ref{fig:variance}.a), and the location of the minimum corresponds to the best-guess values of the exponents, $\nu_{\rm fit}$ and $s_{\rm fit}$. The corresponding horizontal and vertical slices are shown in Figs.~\ref{fig:variance}.b-c). Note that the shape of the reconstructed scaling function $\Lambda_0(\tilde\delta;\nu,s)$ does depend on the trial exponents $\nu$ and $s$. Our best estimate of the scaling function $\Lambda$ shown in Fig.~\ref{fig:nu_v_d}-b) is obtained from $\Lambda_0(\tilde\delta;\nu_{\rm fit},s_{\rm fit})$.\\

\textbf{Bootstrap procedure}
To determine the optimal values of the critical exponents, $\nu_{\rm exp}$ and $s_{\rm exp}$, as well as the confidence intervals, we use a bootstrap method to take into account the impact of the statistical uncertainties of the experimental $\langle p^2 \rangle_t$ data, as well as that of the critical point location $K_c$ \cite{WilliamBook}. Assuming a Gaussian probability distribution for these quantities, and given the standard deviations estimated above, we resample $\langle p^2\rangle_t$ and $K_c$ and repeat the fitting procedure described in the previous paragraph to determine values of the $\nu$ and $s$ exponents. This procedure is repeated $10^4$ times, which provides the ($\nu,s$) samples shown in Fig.~\ref{fig:nu_v_d}-c). We use these samples to determine the best estimates $\nu_{\rm exp}$ and $s_{\rm exp}$ of the critical exponents, as well as the corresponding confidence region (blue ellipse) and the confidence intervals (shown as error bars). Due to the limited amount of experimental data, we estimate that our numerical resampling method provides realistic error estimations. Another option, standard bootstrapping (random sampling of the data with replacement), tends to underestimate confidence intervals in our case due to the small dataset (three to six averages per determination of $\langle p^2 \rangle_t$).
\\

\textbf{Parameters used for testing the two-parameter scale invariance  (Fig.~\ref{fig:scale}).}

\begin{table}[h!]
\centering
\def\arraystretch{1.5}
\begin{tabular}{|c|*6{c|}} \hline
    $\tilde \delta$    & \multicolumn{5}{c|}{$(\delta_K,t)$} \\ \hline
$ -3$ & $(-1.2,69)$ & $(-1.1,103)$ & $(-1,161)$ & $(-0.9,262)$ & $(-0.8,451)$\\
\hline
$ -1.4$ & $(-0.6,50)$ & $(-0.5,117)$ & $(-0.45,190)$ & $(-0.41,293)$ & $(-0.37,470)$\\
\hline
$ +1.4$ 
 & $(+0.6,50)$ & $(+0.5,117)$ & $(+0.45,190)$ & $(+0.41,293)$ & $(+0.37,470)$\\
\hline
$ +3$ & $(+1.2,69)$ &  $(+1.1,103)$ & $(+1,161)$ & $(+0.9,262)$ & $(+0.8,451)$ \\
\hline

\end{tabular}
\end{table}

\end{document}